\documentclass[preprint2]{emulateapj}

\usepackage{natbib}

\shorttitle{X-Ray Measured Dynamics of Tycho's SNR}
\shortauthors{S. Katsuda et al.}

\begin{document}

\title{X-Ray Measured Dynamics of Tycho's Supernova Remnant}

\author{Satoru Katsuda\altaffilmark{1}, Robert
  Petre\altaffilmark{1}, John P. Hughes\altaffilmark{2}, Una
  Hwang\altaffilmark{1,3}, Hiroya Yamaguchi\altaffilmark{4}, Asami
  Hayato\altaffilmark{4}, Koji Mori\altaffilmark{5}, Hiroshi
  Tsunemi\altaffilmark{6}
}

\email{Satoru.Katsuda@nasa.gov, Robert.Petre-1@nasa.gov,
  jackph@physics.rutgers.edu, Una.Hwang-1@nasa.gov,
  hiroya@crab.riken.jp, hayato@crab.riken.jp,
  mori@astro.miyazaki-u.ac.jp, tsunemi@ess.sci.osaka-u.ac.jp
}

\altaffiltext{1}{Code 662, NASA Goddard Space Flight Center,
   Greenbelt, MD 20771, U.S.A.} 

\altaffiltext{2}{Department of Physics and Astronomy, Rutgers
        University, 136 Frelinghuysen Road, Piscataway, NJ 08854-8019,
        U.S.A.}

\altaffiltext{3}{Johns Hopkins University, Baltimore MD 21218, U.S.A.} 

\altaffiltext{4}{RIKEN (The Institute of Physical and Chemical
  Research), 2-1 Hirosawa, Wako, Saitama 351-0198} 

\altaffiltext{5}{Department of Applied Physics, Faculty of Engineering,
University of Miyazaki, 1-1 Gakuen Kibana-dai Nishi, Miyazaki, 889-2192,
Japan}

\altaffiltext{6}{Department of Earth and Space Science, Graduate School
of Science, Osaka University, 1-1 Machikaneyama, Toyonaka, Osaka,
560-0043, Japan}


\begin{abstract}

We present X-ray proper-motion measurements of the forward shock and 
reverse-shocked ejecta in Tycho's supernova remnant, based on three
sets of archival {\it Chandra} data taken in 
2000, 2003, and 2007.  We find that the proper motion of the edge of
the remnant (i.e., the forward shock and protruding ejecta knots)
varies from 0$^{\prime\prime}$.20\,yr$^{-1}$ (expansion index
$m=0.33$, where $R = t^m$) to 0$^{\prime\prime}$.40\,yr$^{-1}$ 
($m=0.65$) with azimuthal angle in 2000-2007 measurements, and
0$^{\prime\prime}$.14\,yr$^{-1}$ ($m=0.26$) to
0$^{\prime\prime}$.40\,yr$^{-1}$ ($m=0.65$) in 2003-2007
measurements.  The azimuthal variation of the proper 
motion and the average expansion index of $\sim$0.5 are consistent
with those derived from radio observations.  We also find proper
motion and expansion index of the reverse-shocked ejecta to be
0$^{\prime\prime}$.21--0$^{\prime\prime}$.31\,yr$^{-1}$ and
0.43--0.64, respectively.  From a comparison of the
measured $m$-value with Type Ia supernova evolutionary models, we
find a pre-shock ambient density around the remnant of
$\lesssim0.2$\,cm$^{-3}$.

\end{abstract}

\section{Introduction}

A supernova (SN) event which Danish astronomer Tycho Brahe
discovered in 1572 is now visible at many wavelengths as Tycho's
SN remnant (SNR).  The SN event has been classified as a normal
Type Ia (i.e., a thermonuclear explosion of a white dwarf star in a close
binary) based on the shape of its light curve (e.g., Baade 1945;
Ruiz-Lapuente 2004) and the X-ray properties (Badenes et al.\
2006).  This is recently confirmed from its light-echo spectrum, 
which reached the earth about 436 yr after Tycho Brahe observed the
direct light from the SN event (Krause et al.\ 2008).  Given the fact
that Type Ia SNe have been used to determine cosmological parameters
such as the Hubble constant and the Dark Energy equation of state
(e.g., Hamuy et al.\ 1996; Riess et al.\ 1998), it is fundamentally
important to understand the explosion mechanism and the progenitor
system.  Tycho's SNR is a very suitable target for such a study.
First, its ejecta-dominated X-ray emission allows us to constrain the
explosion mechanism models as was recently demonstrated by Badenes et
al.\ (2006), who concluded that the explosion producing Tycho's SNR
was most likely a delayed detonation; also, the surviving binary
companion from Tycho's SNR may have been identified (Ruiz-Lapuente et
al.\ 2004; Ihara et al.\ 2007), though the association with the SN
event is still controversial (Kerzendorf et al.\ 2009).  The chemical
abundances of the candidate companion were recently measured to be
consistent with the Galactic trend except for an overabundant Ni/Fe
ratio which suggests contamination by SN ejecta (Hernandez et al.\
2009).  Recent detection of Cr and Mn K$\alpha$ lines in the X-ray
spectrum of the SNR (Tamagawa et al.\ 2009; Yang et al.\ 2009) also
provides an opportunity to constrain the metallicity of the progenitor
star (Badenes, Bravo, \& Hughes 2008).

Another important aspect of Tycho's SNR is the presence of non-thermal
X-ray synchrotron emission produced by accelerated particles (e.g.,
Pravdo \& Smith 1979).  Recent {\it Chandra} images have revealed
that the non-thermal emission is confined to a very thin shell just
behind the forward shock (Hwang et al.\ 2002; Bamba et al.\ 2005;
Warren et al.\ 2005).  The thin filamentary structure of non-thermal
emission is now known to be a common feature in many young SNRs, while
its nature is the subject of considerable theoretical work on
particle acceleration (e.g., Pohl et al.\ 2005; Ballet 2006;
Cassam-Chena\"i et al.\ 2007).  Furthermore, this SNR 
is a possible TeV gamma-ray source, although a firm detection of TeV
gamma-ray emission is not yet available (e.g., V\"olk et al.\ 2008). 

Tycho's SNR potentially provides us with information on SNR evolution,
since the degree of deceleration of the forward shock can be inferred
by combining the current expansion rate with the accurately known age.  
Expansion measurements have been performed at optical, radio, and
X-ray wavelengths.  Optical measurements have focused on the
northwestern (NW), northern (N), and eastern (E) H$\alpha$ filaments.
The expansion index ($m$; $R \varpropto t^m$, where $R$ and $t$ are
the radius and the age of the remnant) for these filaments was
measured to be 0.32--0.41 (Kamper \& van den Bergh 1978), indicating 
significant deceleration  from free expansion ($m = 1$).  The radio
observations revealed that the expansion index varies between 0.2 and
0.8 with azimuthal angle (e.g., Reynoso et al.\ 1997).  The slow
expansion ($m\sim0.2$) seen at the E rim suggests an interaction
between the forward shock and a dense cold cloud.  Such an interaction
was confirmed by the 
subsequent discovery of a dense cloud toward the E rim (Reynoso et
al.\ 1999; Lee et al.\ 2004).  The mean $m$ value was calculated to be
$\sim$0.47, suggesting that overall, this SNR is still evolving toward
the Sedov phase ($m = 0.4$).  Although the mean value of $m$ for radio
observations is higher than that for optical observations, optical and
radio measurements of individual features are consistent with each
other.  In X-rays, the only X-ray expansion measurement to date was
performed using {\it ROSAT} data with moderate angular resolution of
$\sim$8$^{\prime\prime}$ half-power diameter (Hughes 2000).  Interior
features of the remnant show consistent expansion indices between
X-ray and radio observations.  In contrast, the expansion index for
the forward shock measured in X-rays, $m\sim0.71$, is higher than that
for radio.  Since the radio and X-ray images are well correlated with
each other, the expansion rate discrepancy has been puzzling.

Here, we present X-ray proper-motion measurements based on archival
{\it Chandra} data with superior angular resolution of
$\sim$0.5$^{\prime\prime}$ half-power diameter taken in three epochs
between 2000 and 2007.  These {\it Chandra} data allow us to measure
proper motions of the SNR edges as well as the reverse-shocked ejecta
inside the remnant.

\section{Observations and Data Reduction}

{\it Chandra} observed Tycho's SNR in three epochs as summarized in
table~\ref{tab:observations}.  The third epoch comprises two
observations taken on April 23 and 26 in 2007 (ObsID 7639 and 8551;
PI: Hughes).  The exposure time of ObsID 8551 was significantly 
shorter than that of ObsID 7639.  For simplicity, we use only the
longer observation (ObsID 7639) in our analysis.  The information of
the {\it Chandra} observations used in this paper is listed in.  The
time differences between the first and the third observations (i.e.,
2000-2007) and between the second 
and the third (i.e., 2003-2007) are 6.60 yr and 3.99 yr, respectively.
The observations in 2003 and 2007 were specifically intended to allow
proper-motion measurements: both have the same pointing direction,
roll angle, and exposure time. The Advanced CCD Imaging Spectrometer
(ACIS)-I was used for these two observations.  On the other hand, the
2000 observation was not intended for this purpose, and utilized the
back-illuminated ACIS-S3 detector.  However, the 2000 observation is
still useful to measure proper motions.  Therefore, we use the 2000
observation, which in fact provides slightly tighter constraints on
proper motion due to the longer interval between observations.  We
reprocessed all the level-1 event data, applying the standard data
reduction\footnote{http://cxc.harvard.edu/ciao/threads/createL2/} with  
CALDB ver.\ 3.4.0.  The net exposure times for the 2000, 2003, and 2007
observations are 48.9\,ks, 145.7\,ks, and 108.9\,ks, respectively.  In
order to make use of the best position accuracy of the ACIS, we
removed the effects of pixel randomization. 

Figure~\ref{fig:image} shows a three-color {\it Chandra} image
obtained from the 2003 observation.  Red corresponds to low-energy
(0.3--1.0\,keV, mostly Fe L), green to middle-energy (1.0--4.0\,keV,
mostly Si+S+Ar+Ca K), and blue to high-energy (4.0--8.0\,keV, mostly
Fe K blend plus the continuum) bands.  The remnant is outlined by thin
blue (nonthermally-dominated) filaments, while it contains substantial
green (thermally-dominated) fragmentation inside the blue filaments.
The blue filaments are identified as the forward shock (e.g., Warren
et al.\ 2005), while the fragments are the reverse-shocked ejecta
(e.g., Velazquez et al.\ 1998).  We measure proper motion of both the 
forward shock and the shocked ejecta, by comparing
vignetting-corrected 1.0--8.0\,keV images obtained in the three
epoches.

\section{Proper Motion Measurements}

Before measuring proper motions, we try to register the three images 
by aligning the positions of several obvious X-ray point sources
surrounding the remnant.  We determine their positions in each epoch,
by employing {\tt wavdetect} software included in CIAO ver.\ 4.0.  We
detect nine sources around the SNR in the 2003 and 2007 images, while
only five are detected along the NW periphery in the 2000 image.  The
smaller number of sources in the 2000 image is partly due to the fact
that the southern portion of the SNR was not covered by ACIS-S3 chip,
and partly due to the shorter exposure time. If we align only the
sources concentrated in the small NW area, it may cause misalignment
for the rest of the image.  Therefore, we do not correct the
registration between the 2000 and 2007 observations, and instead
incorporate into our proper motion measurement for 2000-2007 a
systematic uncertainty of 0$^{\prime\prime}$.4, which is the 1-sigma
absolute astrometric accuracy for the ACIS camera reported by the {\it
  Chandra} calibration
team\footnote{http://cxc.harvard.edu/cal/ASPECT/celmon/}.  This 
systematic uncertainty seems to be conservative, because the rms
residual calculated for the five point sources detected in both
images is only 0$^{\prime\prime}$.13.  On the other hand, we use the
point sources to register 
images between 2003 and 2007.  Since no optical counterpart was
identified within a 1$^{\prime\prime}$ radius for any of the point sources
based on any catalogs contained in the VizieR catalog
service\footnote{http://vizier.u-strasbg.fr/}, we assume that none have
undergone proper motion between 2003 and 2007.  We
aligned the coordinates of the images so that the point sources have the
same positions, by using {\tt ccmap} and {\tt ccsetwcs} in IRAF
IMCOORDS.  In the alignment procedure, four parameters for the
second-epoch image are allowed to vary: x and y shifts, rotation, and
pixel scale.  The best-fit parameters show differences for both x and
y shifts of less than 0$^{\prime\prime}$.2, no modifications of pixel
scales, and rotation of $3^{\prime}$, respectively.  After the
alignment, the rms residual for the positions of the nine
point sources between 2003 and 2007 images is  
$\sim$0$^{\prime\prime}$.29, which will be taken as the systematic
uncertainty in our proper-motion measurements for 2003-2007.

Figure~\ref{fig:diff} {\it left} and {\it right} show the difference
images in the 1.0--8.0\,keV energy band for 2000-2007 and 2003-2007, 
respectively.  The subtraction is performed after each image was
registered and smoothed by a Gaussian kernel of $\sigma =
0^{\prime\prime}$.492 (i.e., 1 ACIS pixel) which is approximately the
angular resolution of {\it Chandra}'s X-ray telescope.  There are some
artifacts in the difference images: the southern dark (negative)
region seen in the 2000-2007 difference image which is due to the fact
that the 2000 image has no exposure there, the structures running from
NW to SE and from NE to SW caused by the ACIS chip gaps, and stripes
running from NE to SW caused by bad columns.  Apart from these
artifacts, we can clearly see motion of the X-ray features as 
adjacent black and white features for the forward shock (the edge of
the remnant) and the reverse-shocked ejecta (knotty features inside
the remnant).  We quantify the proper motion of these features in the
following sections.

\subsection{SNR Edges}
\subsubsection{Forward Shock}

In order to measure proper motions accurately for the forward shock,
we generate one-dimensional X-ray profiles along the apparent
direction of shock motions, following Katsuda, Tsunemi, \& Mori
(2008).  These one-dimensional X-ray profiles are extracted from the
39 small boxes (25$^{\prime\prime}\times$25$^{\prime\prime}$)
indicated in Fig.~\ref{fig:image}.  The position angle of each box is
chosen such that it is as perpendicular as possible to the shock front
(by eye).  The choice of position angle does not affect the proper
motion measurement within at least $\pm6$ degrees difference, which we
checked for several boxes.  In order to include the calibration
uncertainty of the ACIS effective area, we add a
5\footnote{http://cxc.harvard.edu/cal/} systematic error term in 
quadrature to the statistical error for each data point.  Each profile
is binned by 0$^{\prime\prime}$.5 along the radial direction of the
SNR.  An example profile is shown in Fig.~\ref{fig:onedim}, in which
the shift due the motion of the shock is readily apparent.  Shifting
the first-epoch profile and calculating the $\chi^2$ values from the
difference between the two profiles at each shift position, we search
for the best-matched shift position for each box.
Figure~\ref{fig:chi} shows the $\chi^2$ distribution as a 
function of shift position for the example profiles in
Fig.~\ref{fig:onedim}.  By applying a quadratic function for  
the $\chi^2$ profile, we measure the best-shift position where the
minimum $\chi^2$ value occurs.  In this way, we obtain two (2000-2007
and 2003-2007) results for each box.  The reduced-$\chi^2$ value,
which is nearly equal to $\chi^2$ value divided by the number of bins
in each box (here 36), falls in the ranges 0.9--10.4 for 2000-2007 and
0.9--2.6 for 2003-2007 at the best-matched shift positions.  Two
neighboring boxes (azimuth angles = 78$^\circ$ and 85$^\circ$
counterclockwise from N) for the 2000-2007 measurements have large
($>$3.7) reduced-$\chi^2$ values.  This is partly due to the fact that
these two include the brightest parts among the entire SNR edge,
resulting in smaller statistical errors than those of the others, and
partly due to the fact that the angular resolution around the two
boxes are significantly different between 2000 and 2007 observations.
At the location of the two boxes, we simulate point spread functions
(PSFs) for 2000 and 2007 images by using 
{\tt chart}\footnote{http://cxc.harvard.edu/chart/} and 
{\tt marx}\footnote{http://cxc.harvard.edu/chart/threads/marx/} 
software, and find that the shape of the PSF for the 2000 image is 
elongated in the direction of shock motion, while that for the 2007
image is not.  If we fit the PSFs with a Gaussian kernel, we obtain
$\sigma \sim 1^{\prime\prime}.7$ and $\sigma \sim 0^{\prime\prime}.9$
for the 2000 and 2007 images, respectively.  Therefore, we smooth the
2007 image by a Gaussian kernel of $\sigma = 1^{\prime\prime}.476$
(i.e., 3 ACIS pixels), such that it has the same image resolution as
that of the 2000 image.  In this way, we obtained reasonable fits for
all the box regions (reduced-$\chi^2 <$ 2.6).

Also notable in Fig.~\ref{fig:onedim} is possible flux variations
among the three epochs.  We find that X-ray flux variations among the
three epochs are within $\sim$10\% for all the boxes.  The 10\% flux
changes are larger than the ACIS effective area calibration
uncertainty (5\%).  Considering that the magnetic field strength in
Tycho's SNR is thought to be in the milligauss range (e.g., 0.1--10
mG, Reynolds \& Ellison 1992; 0.03--0.4 mG, Warren et al.\ 2005),
year-scale flux variations of the synchrotron emission seen in
RXJ11713.7$-$3946 (Uchiyama et al.\ 2007) or Cas~A SNRs (Uchiyama \&
Aharonian 2008) may be expected in Tycho's SNR as well.  However, the
possible flux variations in Tycho's SNR are small compared with those
seen in these other SNRs, where some knots showed more than 100\% flux
variations. Further detailed investigations are needed to determine
whether the flux variation is due to the flux change of the
synchrotron emission, and is beyond the scope of this paper.

\subsubsection{Ejecta Knots Protruding beyond the Forward Shock}

\begin{figure}
\includegraphics[angle=0,scale=0.45]{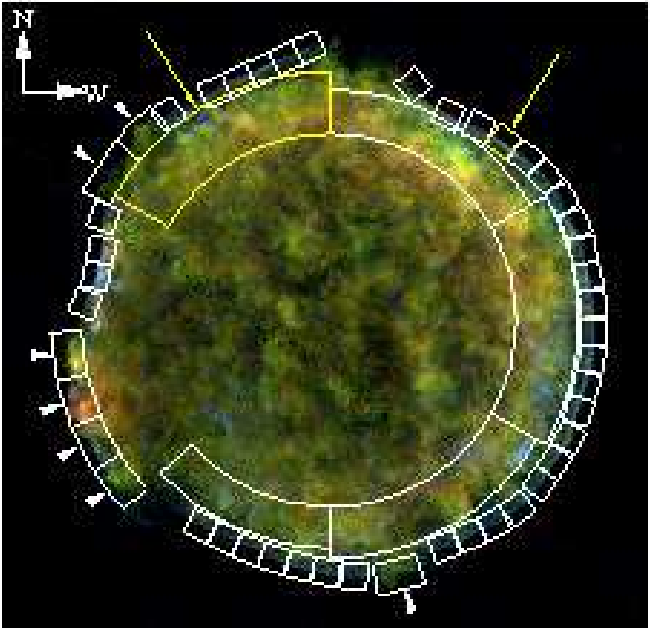}\hspace{1cm}
\caption{{\it Chandra} three-color image of Tycho's SNR after
  vignetting effects are corrected.  Red, green, and blue represent
  0.3--1.0\,keV, 1.0--4.0\,keV, and 4.0--8.0\,keV X-rays,
  respectively.  The image is 
  binned by 0$^{\prime\prime}$.492 and has been smoothed by a Gaussian
  kernel of $\sigma = 0^{\prime\prime}.984$.  The intensity scale is
  square root.  The small boxes, small annular pieces with triangular
  marks, and a large annulus divided by five sectors are where we
  investigate the proper motion of the forward shocks, ejecta knots,
  and reverse-shocked ejecta, respectively.  We show an example
  one-dimensional X-ray profile from the yellow box indicated by 
  an arrow in Fig.~\ref{fig:onedim}.  Example $\chi^2$ or $C$
  values calculated  for the yellow box and the yellow annular sector
  with an arrow are shown in Fig.~\ref{fig:chi} and Fig.~\ref{fig:C}. 
} 
\label{fig:image}
\end{figure}

\begin{figure}
\includegraphics[angle=0,scale=0.45]{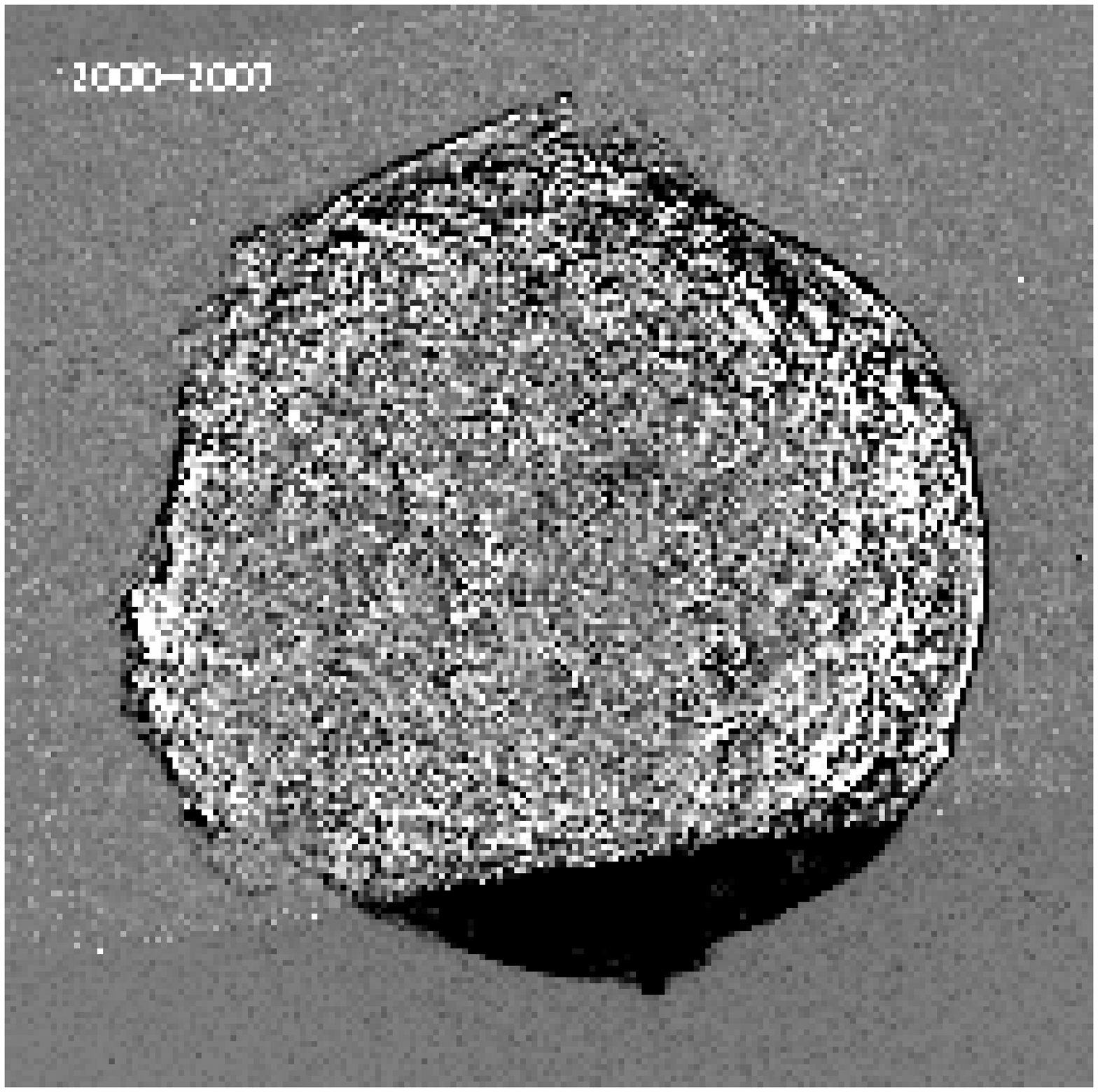}
\includegraphics[angle=0,scale=0.45]{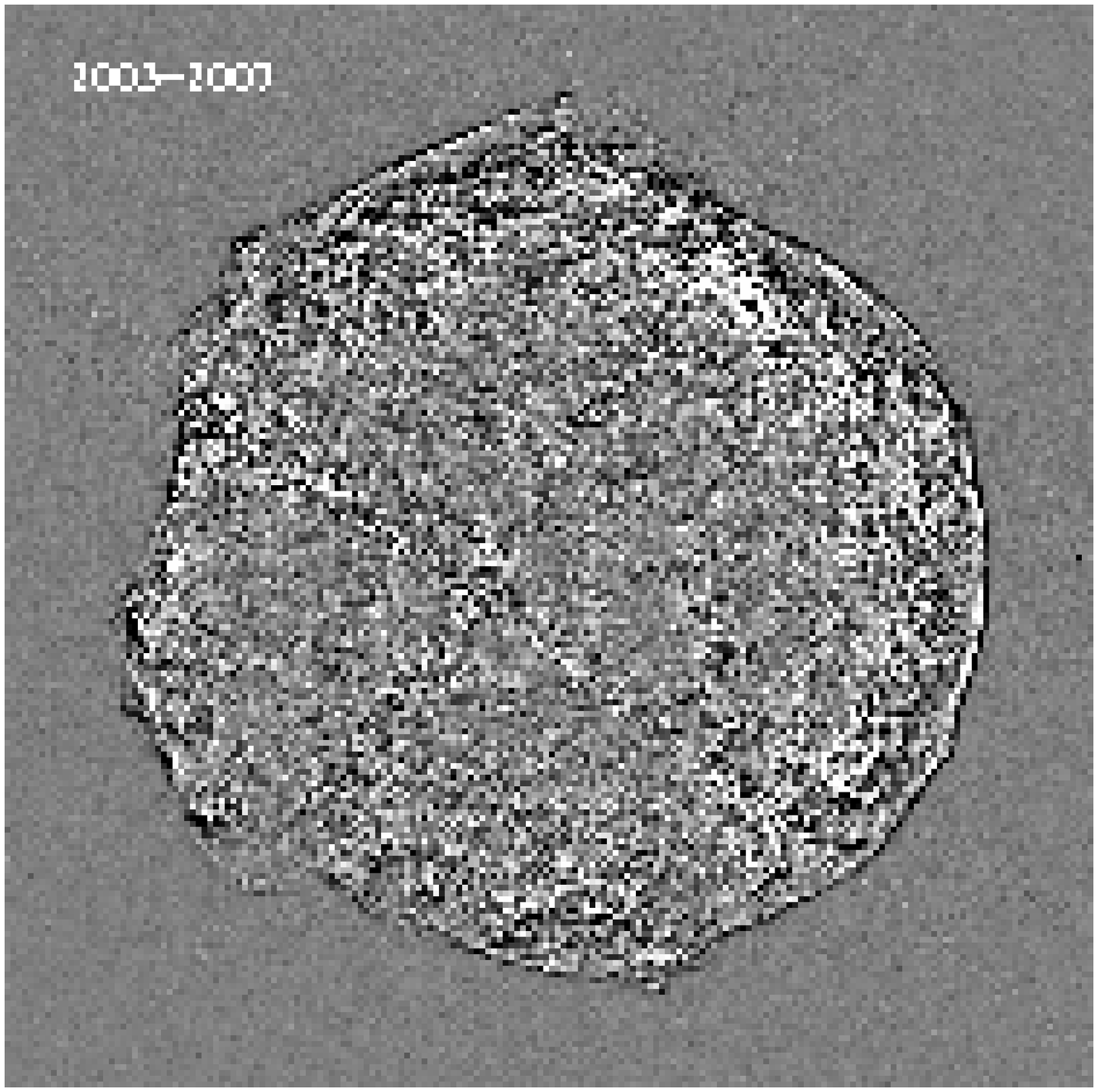}
\caption{{\it Left}: Difference image between 2000 and 2007
  observations in 1.0--8.0\,keV band.  The original images before
  subtraction are binned by 0$^{\prime\prime}$.492 and smoothed
  using a Gaussian kernel of $\sigma = 0^{\prime\prime}.492$.  The
  intensity is linearly scaled from -3 to +3 counts\,pixel$^{-1}$.
  The southern dark (negative) region of the remnant is due to the
  fact that the observation in 2000 does not cover there.  {\it
  Right}: Same as {\it left} but for 2003-2007.  The structures
  running from NW to SE and from NE to SW are caused by the gaps of
  ACIS chips.  Also visible artifacts are stripes running from NE to
  SW caused by bad columns.  
}
\label{fig:diff}
\end{figure}

\begin{figure}
\includegraphics[angle=0,scale=0.35]{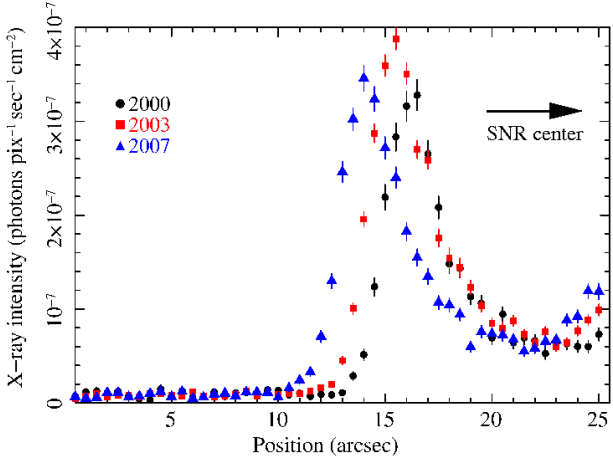}
\caption{Example one-dimensional profiles extracted from the yellow box
  shown in Fig.~\ref{fig:image}.  Data points with black circles, 
  red boxes, and blue triangles represent the 2000, 2003,
  and 2007 epochs, respectively.}  
\label{fig:onedim}
\end{figure}

\begin{figure}
\includegraphics[angle=0,scale=0.35]{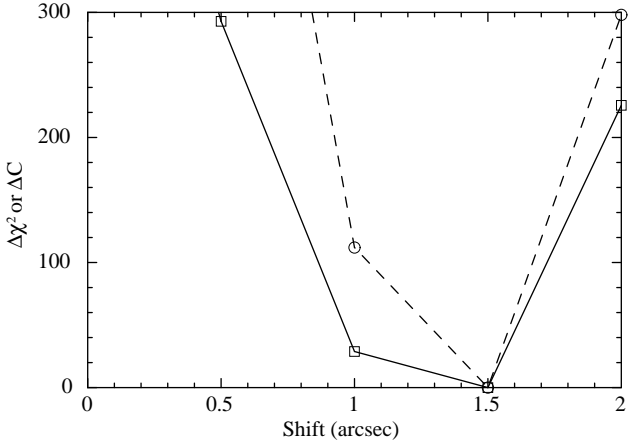}
\caption{Example $\chi^2$- and $C$-value distributions as a function
  of shift position.  The $\chi^2$ values with a solid line
  are calculated from the 2003 and 2007 one-dimensional profiles shown
  in Fig.~\ref{fig:onedim} (see, section 3.1.1), while the $C$ values
  with a dashed line are calculated from 2003 and 2007
  polar-coordinate images for the same region (see, section 3.1.2).} 
\label{fig:chi}
\end{figure}

There are breakout regions along the periphery of the SNR, as
indicated by the small annular segments with triangular marks in
Fig.~\ref{fig:image}.  In these regions, the SNR edges show fragmented
features which look like Rayleigh-Taylor fingers.  It is 
already shown that this fragmentation is formed by ejecta knots
protruding beyond the forward shock, based on spectral analysis of
{\it XMM-Newton} data (Decourchelle et al.\ 2001) as well as principal
component analysis of {\it Chandra} data (Warren et al.\ 2005).  Since 
these ejecta knots show completely different morphology from
the filamentary structure of the forward shock, we cannot define box
regions relative to the shock front.  We thus use two-dimensional
images instead of one-dimensional profiles to measure the proper
motion.  We follow the method employed by DeLaney \& Rudnick 
(2003), who measured proper motion of the forward shock in Cas~A.  We
first generate polar-coordinate images for the three observations.
Then, radially shifting the polar-coordinate image from 2000 or 2003
by steps of 0$^{\prime\prime}$.5 with respect to 2007 image, 
we search for the best-matched shift positions.  To quantitatively obtain the
best-shift position, we calculated the $C$ value at each shift position,
where $C$ is expressed as (Cash 1979; Vink 2008), \[C =
-2\mathrm{ln}P = -2\sum_{i}(n_i\mathrm{ln}m_i-m_i-\mathrm{ln}n_i!),\] 
with $n_i$ being the observed counts in a pixel $i$ and $m_i$ being
the model counts in the same pixel.  We here assume that the 2007 
observation is the model image.  Using the $C$ value rather than
$\chi^2$ is necessary when the number of counts is so low
that the shape of the Poisson distribution becomes significantly different
from that of the Gaussian distribution.  In fact, this is the case for
a number of pixels in the data used here.  Since the analyzed areas
(i.e., the small annular segments in the Fig.~\ref{fig:image}) include
regions outside the remnant, they contain a significant number of pixels 
with zero counts.  Although we
cannot sum pixels with $n_i$ (or $m_i$) = 0 according to the above
equation, we have to sum the same numbers of pixels for each shift
position.  To this end, we assume the observed and model counts in the
pixels having zero counts to be unity.  We check the validity
of this assumption, by applying the two-dimensional method for several
forward shock regions and confirming that the proper motion value obtained
from the one- and two-dimensional methods are consistent with each
other.  As an example, the $C$ values calculated for the 
yellow box in Fig.~\ref{fig:image} is shown as a
function of shift position in Fig.~\ref{fig:chi} (with dashed lines),
from which we can see that the best-shift position from
the two-dimensional method ($C$ value) is the same as that from the
one-dimensional method ($\chi^2$ value).  The best-shift position for
each area is estimated by fitting a quadratic function for each
profile.  In estimating uncertainties, $\Delta C$ (= $C_\mathrm{min}
- C$) can be treated as $\Delta \chi^2$ (Cash 1979).  In this way, we
obtained two sets (2000-2007 and 2003-2007) of proper motion
measurements for each small annular segment.

\subsubsection{Results for the SNR Edge}

Table~\ref{tab:results1} gives the proper motion, expansion rate,
and expansion index (which are derived from ``the expansion rate''
$\times$ ``the age of the remnant, i.e., 430 or 432 yr for 2000-2007
or 2003-2007, respectively'') for all areas including the forward
shock (small boxes in Fig.~\ref{fig:image}) and ejecta knots (small
annular pieces in Fig.~\ref{fig:image}).  We also list the mean values 
which are simply the sum for all areas (both 2000-2007 and 
2003-2007 measurements) divided by the number of areas.  To 
calculate the expansion rate and expansion index, we need to
determine the expansion center.  We take the geometric center (J2000)
of RA$=$00$^\mathrm{h}$25$^\mathrm{m}$19$^\mathrm{s}$.4, DEC
$=+64^{\circ}08^{\prime}13^{\prime\prime}.98$ determined by minimizing 
the ellipticity of the forward shock (Warren et al.\ 2005).  
Figure~\ref{fig:prop_fs} plots the values as a function of
azimuthal angle.  Black and red points represent 
2000-2007 and 2003-2007 measurements, respectively.  Data
points with filled boxes are responsible for the forward shock 
and open circles for the ejecta knots.  The errors quoted in 
Fig.~\ref{fig:prop_fs} represent statistical 1-sigma uncertainties.
The magnitude of the systematic uncertainties are indicated in the
figure as black and red bars, and are generally $\sim$3 times larger
than the statistical uncertainties.  If both statistical and
systematic uncertainties are considered, the two results for each area
are all consistent with each other except for the E rim
(70$^\circ$--90$^\circ$ in azimuthal angle) where we see indications 
of significant shock deceleration.  The variation of the proper
motion ranges from 0$^{\prime\prime}$.14\,yr$^{-1}$ at the E rim to 
0$^{\prime\prime}$.40\,yr$^{-1}$ at the western (W) rim in 2003-2007 
measurements, but is relatively more modest variation at
0$^{\prime\prime}$.20\,yr$^{-1}$ to 0$^{\prime\prime}$.40\,yr$^{-1}$
in 2000-2007 measurements.

\subsection{Reverse-Shocked Ejecta}

We now describe our measurement of the proper motion of the
reverse-shocked ejecta.  Warren et al.\ (2005) located the position of
the reverse shock at $\sim$3$^{\prime}$ radius, which is the peak of
the radial profile of the Fe~K shell line.  They also reported that
the contact discontinuity between the reverse-shocked ejecta and the
swept-up interstellar medium (ISM) is located around
3$^{\prime}$.67--4$^{\prime}$.5 radius.  We thus focus on an annular
region between the reverse shock (at 3$^{\prime}$ radius) and the
contact discontinuity (at 3$^{\prime}$.71--4$^{\prime}$ radius
depending on azimuthal angle, based on Fig.~4 in Warren et al.\
2005).  To investigate azimuthal variations, we divide the annular
region into five sectors as shown in Fig.~\ref{fig:image}.  The five
sectors represent NE, southeast (SE), southwest (SW), W, and
NW defined as: 
NE (0$^\circ$--60$^\circ$ in azimuthal angle), 
SE (132$^\circ$--180$^\circ$), 
SW (180$^\circ$--240$^\circ$), 
W (240$^\circ$--300$^\circ$), 
NW (300$^\circ$--360$^\circ$).
Note that we exclude the E sector from our analysis.  In these
regions, Warren et al.\ (2005) could not identify the reverse shock
due to the very weak Fe~K line emission there, nor do we see clear
evidence of ejecta motion.

As mentioned in Sec.~2, the emission from the reverse-shocked
ejecta appears as fragmented structures and is completely different
from the filamentary structures of the forward shock.  Therefore, we used 
the two-dimensional method used for measuring the proper motion of
the ejecta knots (see Sec.~3.1.2).  The $C$-value calculated for the 
example sector shown in yellow in Fig.~\ref{fig:image} is
plotted as a function of shift position in Fig.~\ref{fig:C}.
Table~\ref{tab:results2} lists the proper motion, expansion rate, and
expansion index for individual regions with their statistical
uncertainties as well as their mean values.  Figure~\ref{fig:prop_ej}
shows the results as a function of azimuthal angle.  Black and red
points represent 2000-2007 and 2003-2007 measurements, respectively.
Errors with the best-fit values represent only statistical
uncertainties, while the systematic errors are shown as black and red
bars.  The azimuthal variations for the proper motion range from
0$^{\prime\prime}$.21\,yr$^{-1}$ at the NW sector to
0$^{\prime\prime}$.31\,yr$^{-1}$ at the SW sector, although they
do not show significant variation within their combined statistical
and systematic uncertainties.

It is of interest to investigate the X-ray energy dependency
of the expansion, especially for the reverse-shocked ejecta whose
emission lines show distinct radial distributions: the radial
profiles of Si~K and S~K line emission peak outside that of the
Fe~K line emission (Hwang \& Gotthelf 1997; Decourchelle et al.\ 2001). 
Furthermore, Doppler velocities of the Si~K and S~K lines were recently
reported to be larger than that of Fe~K line (Furuzawa et al.\ 2009;
Hayato et al.\ in preparation), indicating higher velocity of the Si
and S ejecta than of Fe.  Therefore, we performed expansion
measurements of the shocked ejecta for selected energy bands.
Unfortunately, we could not derive meaningful results for a narrow
energy band containing the Fe~K line due to insufficient statistics.
On the other hand, we find that results for the energy bands of
0.75--1.43\,keV (dominated by the Fe L lines) and 1.63--4.1\,keV
(dominated by Si and S K lines), which are not shown here, are both
consistent with those derived from the 1.0--8.0\,keV band, as we
expect given that images in these energy bands are very similar (Hwang
\& Gotthelf 1997; Decourchelle et al.\ 2001).

\section{Discussion}

\subsection{Comparison among Radio, Optical, and X-Ray Measurements}

The proper motion along the edge of the SNR (i.e., forward shock and
ejecta knots) varies with azimuthal angle.  In general, the variation
is quantitatively consistent with radio measurements in which slower
motions are observed at the E rim and faster motions at the W rim 
(0$^{\prime\prime}$.17\,yr$^{-1}$--0$^{\prime\prime}$.31\,yr$^{-1}$ 
by Strom et al.\ 1982;
0$^{\prime\prime}$.14\,yr$^{-1}$--0$^{\prime\prime}$.42\,yr$^{-1}$ by
Reynoso et al.\ 1997), Our results are also 
consistent with the optical measurements of the NW, N, and E rims
(Kamper \& van den Bergh 1978).  The averaged expansion index for all
areas is $\sim$0.52.  This value is consistent with or close to those
determined by radio observations: 0.47$\pm$0.07 
(Strom et al.\ 1982), 0.462$\pm$0.024 (Tan \& Gull 1985),
0.471$\pm$0.028 (Reynoso et al.\ 1997).



\subsection{Interpretation of the Azimuthal Variation}

The shock motion in the E rim is the slowest along the entire rim.
This agrees with the radio proper motion measurements and further
supports the idea that the forward shock is interacting with a dense
cold cloud in this region (Strom 1983; Reynoso et al.\ 1999; Lee et
al.\ 2004).  The interaction is suggested to be very recent (within
the last 50\,yr) and strong (the cloud density of 160--325\,cm$^{-3}$)
from radio observations (Reynoso et al.\ 1999).  The X-ray data show a
hint of deceleration along eastern rim (especially at
70$^\circ$--90$^\circ$ in azimuthal angle).  This possible 
deceleration, if true, would strongly support the view that the
Tycho's SNR has been recently interacting with the dense cloud along
the E rim.

\begin{figure}
\includegraphics[angle=0,scale=0.7]{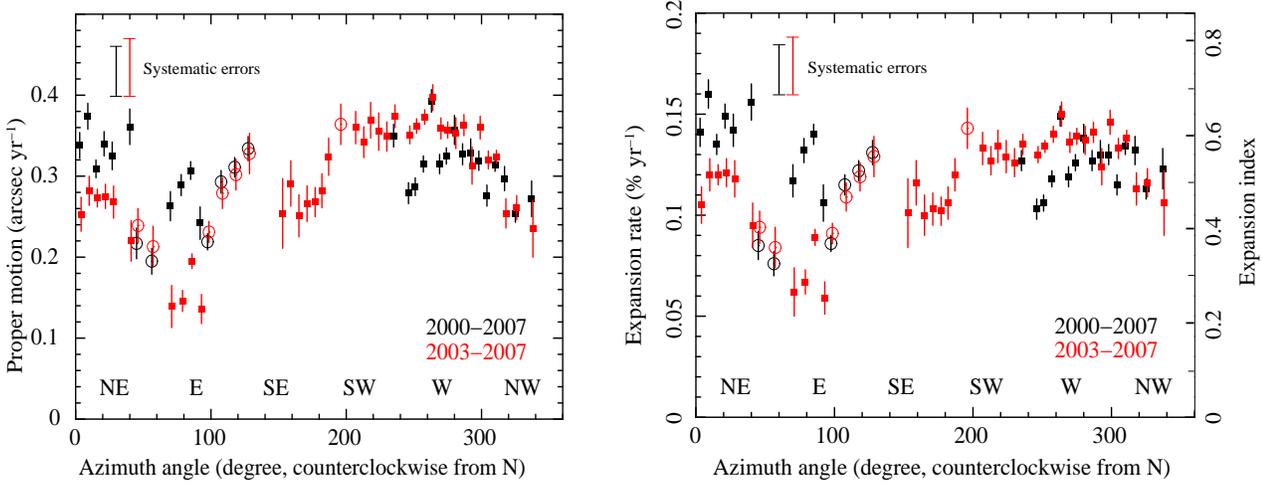}
\caption{{\it Left}: Proper motion for the forward shocks and the
  ejecta knots as a function of azimuthal angle (counterclockwise from
  the N).  Black and red are derived from 2000-2007 and 2003-2007
  measurements, respectively.  Data points with filled boxes and open
  circles represent the forward shocks and the ejecta knots,
  respectively.  The data for 2003-2007 measurements are displaced by
  1 degree in azimuthal angle for clarity.  The southern portion of
  the remnant (152$^\circ$--229$^\circ$ in azimuth angle) was not
  covered by the 2000 observation.  {\it Right}: Same as left but for
  expansion rate (left axis) and expansion index (right axis).} 
\label{fig:prop_fs}
\end{figure}

\begin{figure}
\includegraphics[angle=0,scale=0.35]{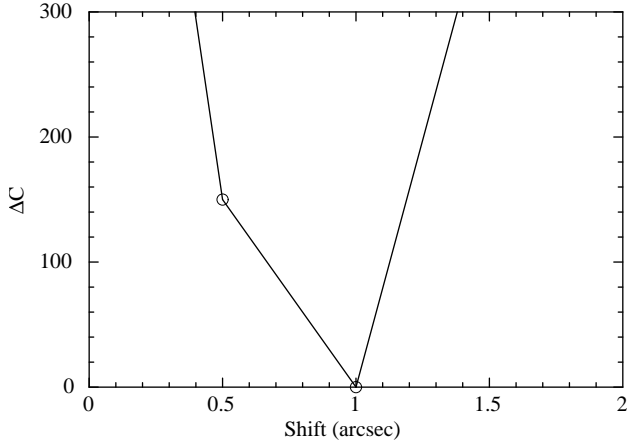}
\caption{Example $C$-value distribution as a function of shift
  position for the reverse-shocked ejecta from the yellow (NE) annular
  sector in Fig.~\ref{fig:image} calculated from 2003 and 2007 polar
  coordinated images.}  
\label{fig:C}
\end{figure}

\begin{figure}
\includegraphics[angle=0,scale=0.7]{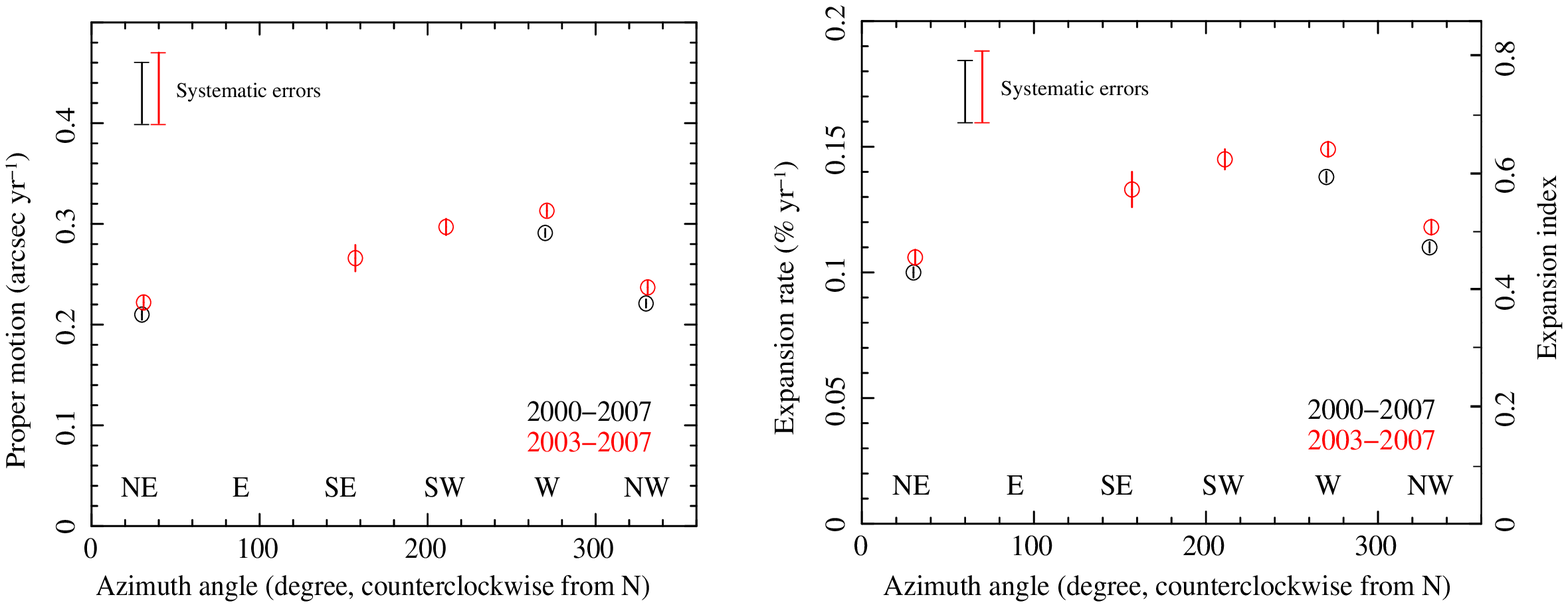}
\caption{Same as Fig.~\ref{fig:prop_fs} but for the reverse-shocked ejecta.}
\label{fig:prop_ej}
\end{figure}

In Fig.~\ref{fig:prop_fs} {\it left}, we also see a possible
variation of the proper motion of the forward shock along the 
W--NW rim.  The value of proper motion is generally proportional to
the shock radius in this region (see, Fig.~\ref{fig:prop_fs} or Fig.~4
in Warren et al.\ 2005), resulting in fairly a constant expansion rate
throughout the region as shown in Fig.~\ref{fig:prop_fs} {\it right}.  The
gradual increase of the proper motion from the NW rim to the W rim
suggests that the ambient density along the NW rim is denser than
that along the W rim.  Since there is no evidence in the radio of a
dense cloud around the NW rim (Reynoso et al.\ 1999; Lee et al.\
2004), a recent strong shock deceleration does not 
seem to be the case.  Therefore, we suggest that the forward shock
has been constantly decelerated in these regions.  The slower shock
motion as well as the smaller shock radius in the NW direction most
likely reflect a slightly denser ambient medium in this direction than
that in the W direction.  We can safely assume pressure
equilibrium behind the forward shock (i.e., $n_0 v_\mathrm{s}^2$,
where $n_0$ is the ambient density and $v_\mathrm{s}$ is the shock
velocity) along the NW rim to the W rim.  Using the proper
motions of $\sim$$0^{\prime\prime}.26$\,yr$^{-1}$ at the NW rim and
$\sim$$0^{\prime\prime}.34$\,yr$^{-1}$ at the W rim, we find that the
ambient density ratio of NW rim to W rim is $\sim$2.  The higher
ambient density toward the NW direction should yield a stronger (i.e.,
slower) reverse shock in this region.  This is indeed consistent with
the brighter X-ray intensity of the ejecta in the NW sector than in
the W sector (see Fig.~\ref{fig:image}) and the possibility that the
reverse-shocked ejecta may be moving slower in the NW sector than in
the W sector (though this is not significant within the systematic
uncertainties).

\subsection{Ambient Density}

We estimate the ambient density around Tycho's SNR, based on a
comparison of the measured expansion index with theoretical SNR
evolutionary models.  Dwarkadas \& Chevalier (1998) used one-dimensional
simulations to investigate SNR evolution for power-law and exponential
ejecta density profiles.  They predicted that the exponential profile
shows a density curve increasing from the reverse shock to the contact
discontinuity, while the power-law profile shows a decrease.  The
density profile expected for the exponential profile qualitatively
matches that observed in Tycho's SNR (Hwang \& Gotthelf 1997).
Furthermore, the exponential profile is a better approximation than
the power-law profile for the SN ejecta in detailed explosion models.
Dwarkadas \& Chevalier (1998) thus concluded that the exponential
profile is more suitable for Tycho's SNR than the power-law profile.
Dwarkadas (2000) extended the evolutionary model with the exponential
ejecta density profile to two dimensions, including the effect of
Rayleigh-Taylor fingers acting at the contact discontinuity.  We
compare our measured expansion index for the forward shock (i.e., SNR
edge) with that expected from the two-dimensional model examined in a
uniform ambient medium. In this model, the evolution of the expansion
index of the forward shock can be scaled by a characteristic time, 
248$E_{51}^{-1/2}$($M_\mathrm{ej}/M_\mathrm{Ch}$)$^{5/6}$$n_0^{-1/3}$ 
yr (e.g., Dwarkadas \& Chevalier 1998), where $E_{51}$ is the
explosion energy in units of 10$^{51}$\,ergs, $M_\mathrm{Ch}$ is
defined to be 1.4\,$M_\odot$, and $n_0$ is the ambient H density in 
units of cm$^{-3}$.  We exclude the E rim where
the cloud-shock interaction is obvious (60$^{\circ}$--100$^{\circ}$ in 
azimuthal angle), since the model is calculated assuming spherical
symmetry.  Thus, we use an average $m$-value for the SNR edge of
$\sim$0.54.  This $m$-value requires an age equal to the
characteristic time according to Fig.~1 in Dwarkadas (2000).  For
the age of Tycho's SNR (432\,yr), we derive the ambient density to be 
$\sim$0.2$E_{51}^{-3/2}$($M_\mathrm{ej}/M_\mathrm{Ch}$)$^{5/2}$\,cm$^{-3}$.  
This density is at least three times lower than that required from
a hydrodynamic model to reproduce structures of the ionization age and
the electron temperature seen in the X-ray line emission (Dwarkadas \&
Chevalier 1998).  Also, the density is about five times lower than
that preferred by detailed comparison between hydrodynamical Type Ia
SN models and observed line emission in X-ray spectra (Badenes et al.\
2006).  On the other hand, this density is well within 
the upper limit for the ambient density of 0.3\,cm$^{-3}$ (with an
uncertainty of factor $\sim2$) based on the contribution of the
thermal emission from the shocked ambient medium in the {\it Chandra}
spectra (Cassam-Chena\"i et al.\ 2007).  It is also consistent with
both the previous optical measurement of 0.3\,cm$^{-3}$ for the NE
filament (Kirshner, Winkler, \& Chevalier 1987) and the upper limit of 
0.4\,cm$^{-3}$ based on the gamma-ray flux (V\"olk et al.\ 2008).
Therefore, comparisons of hydrodynamical models with the X-ray
line emission agree in the need for higher densities than the other
estimates.  As Dwarkadas \& Chevalier (1998) noted, it might be worth
considering a more complex circumstellar and/or ejecta structure in
the hydrodynamical model to explain the X-ray line emission.

We note that the effects of efficient particle acceleration are not 
considered by the models above, but are now believed to influence
the forward shock in this SNR (Warren et al.\ 2005; Cassam-Chena\"i et
al.\ 2007).  Efficient particle acceleration would yield high
compression ratios behind the shock well in excess of 4, resulting in
stronger shock deceleration than if efficient particle acceleration
were absent.

There are a number of authors who considered particle acceleration
effects on the SNR evolution (e.g., Decourchelle, Ellison, \& Ballet 
2000; Ellison et al.\ 2007).  Ellison et al.\ (2007) showed in Fig.~2
in their paper that the model with $\epsilon_\mathrm{rel} \backsimeq
63$\% roughly matches the radius ratios among the forward and reverse
shocks, and the contact discontinuity observed in Tycho's SNR, where 
$\epsilon_\mathrm{rel}$ is defined as the percentage of energy flux
crossing the shock that ends up in relativistic particles.  This model
predicts that the radius and velocity of the forward shock expected
with no relativistic particles ($\epsilon_\mathrm{rel} \backsimeq
0$\%) are $\sim$1.1 and $\sim$1.3 times larger than those observed
($\epsilon_\mathrm{rel} \backsimeq 63$\%).  This means that the
observed $m$-value would be modified to 0.65 (= 0.54$\times$1.3/1.1)
when compared with the SNR evolutionary models without relativistic
particles.  By using the $m$-value of 0.65, we derive the time to be
0.2 in characteristic units again based on Fig.~1 in Dwarkadas (2000).
Then, the ambient density is calculated to be 
$\sim$0.0015\,$E_{51}^{-3/2}$($M_\mathrm{ej}/M_\mathrm{Ch}$)$^{5/2}$\,cm$^{-3}$.
Since this model requires even lower ambient density than that from
the model without particle acceleration ($\sim$0.2\,cm$^{-3}$), we
conclude that the ambient density around Tycho's SNR is likely less
than $\sim$0.2\,cm$^{-3}$.  Further detailed SNR evolutionary models
considering the effect of relativistic particles would be helpful in
deriving an accurate ambient density.

\section{Conclusion}

We have presented X-ray proper motion measurements of the edge
(i.e., forward shock and ejecta knots) and the reverse-shocked ejecta
of Tycho's SNR.  The azimuthal variation of the proper motions
along the edge as well as the average expansion index of $\sim$0.5 are 
consistent with those derived from the most recent radio measurements
(Reynoso et al.\ 1997).  The proper motions for the
reverse-shocked ejecta is measured to be
0$^{\prime\prime}$.21--0$^{\prime\prime}$.31\,yr$^{-1}$.  Comparison 
of the expansion index of the forward shock and the reverse-shocked
ejecta with an evolutionary model of a Type Ia SN gave us a typical
pre-shock ambient density of less than $\sim$0.2\,cm$^{-3}$.

\acknowledgments

We thank the anonymous referee for thorough reading and critical 
comments which significantly improved the quality of this paper.
S.K.\ is supported by a JSPS Research Fellowship for Young
Scientists. S.K.\ is also supported in part by the NASA grant under
the contract NNG06EO90A.  J.P.H. acknowledges partial support from
Chandra grant number GO7-8071X to Rutgers University.

\begin{deluxetable}{lccccc}
\tabletypesize{\tiny}

\tablecaption{{\it Chandra} observations used in this paper}
\tablewidth{0pt}
\tablehead{
\colhead{ObsID}&\colhead{ObsDate}
&\colhead{Array}&\colhead{Coordinate (J2000)} &\colhead{Exposure Time
  (ksec)} 
}
\startdata
115&2000-9-20&ACIS-S&00$^{\rm h}$25$^{\rm m}$07$^{\rm s}$.34, 64$^\circ$09$^{\prime}$42$^{\prime\prime}$.7& 48.9\\
3837&2003-4-29&ACIS-I&00$^{\rm h}$25$^{\rm m}$21$^{\rm s}$.48, 64$^\circ$08$^{\prime}$13$^{\prime\prime}$.9& 145.7\\
7639&2007-4-23&ACIS-I&00$^{\rm h}$25$^{\rm m}$22$^{\rm s}$.02, 64$^\circ$08$^{\prime}$13$^{\prime\prime}$.1& 108.9
\enddata
\label{tab:observations}
\end{deluxetable}

\begin{deluxetable}{lccccc}
\tabletypesize{\tiny}

\tablecaption{Summary of proper-motion measurements for forward
  shock (FS) and ejecta knots (EK)}
\tablewidth{0pt}
\tablehead{
\colhead{Azimuth (degree)}&\colhead{Proper
  motion (arcsec\,yr$^{-1}$)}&\colhead{Expansion rate
  (\%\,yr$^{-1}$)}&\colhead{Expansion index}  
}
\startdata
&2000-2007 | 2003-2007&2000-2007 | 2003-2007&2000-2007 | 2003-2007
\\
\hline
3 (FS) & $0.338\pm0.016$ | $0.256\pm0.024$
& $0.141\pm0.007$ | $0.107\pm0.010$
& $0.605\pm0.029$ | $0.461\pm0.043$ \\
9 (FS) & $0.374\pm0.016$ | $0.279\pm0.021$ 
& $0.160\pm0.007$ | $0.119\pm0.009$
& $0.687\pm0.029$ | $0.516\pm0.040$ \\
15 (FS) & $0.309\pm0.011$ | $0.270\pm0.015$
& $0.135\pm0.005$ | $0.119\pm0.007$ 
& $0.582\pm0.020$ | $0.512\pm0.028$ \\
21 (FS) & $0.340\pm0.015$  | $0.273\pm0.019$ 
& $0.149\pm0.007$ | $0.120\pm0.008$ 
& $0.641\pm0.028$ | $0.517\pm0.036$ \\
27 (FS) & $0.325\pm0.017$ |  $0.261\pm0.024$
& $0.142\pm0.008$ | $0.114\pm0.011$
& $0.612\pm0.033$ | $0.495\pm0.046$ \\
40 (FS) & $0.361\pm0.022$ | $0.217\pm0.029$
& $0.156\pm0.010$ | $0.094\pm0.013$
& $0.672\pm0.041$ | $0.407\pm0.055$ \\
45 (EK) & $0.217\pm0.019$ | $0.239\pm0.021$
& $0.085\pm0.007$ | $0.094\pm0.008$
& $0.366\pm0.032$ | $0.405\pm0.036$ \\
56.25 (EK) & $0.195\pm0.016$ | $0.213\pm0.025$
& $0.076\pm0.006$ | $0.084\pm0.010$
& $0.328\pm0.027$ | $0.361\pm0.042$ \\
70 (FS) & $0.263\pm0.018$ | $0.142\pm0.029$
& $0.117\pm0.008$ | $0.063\pm0.013$
& $0.503\pm0.034$ | $0.272\pm0.055$ \\
78 (FS) &  $0.289\pm0.013$ | $0.173\pm0.012$
& $0.103\pm0.006$ | $0.079\pm0.006$ 
& $0.442\pm0.025$ | $0.341\pm0.024$ \\
85 (FS) & $0.307\pm0.011$ | $0.204\pm0.012$
& $0.108\pm0.005$ | $0.093\pm0.005$
& $0.466\pm0.021$ | $0.402\pm0.024$ \\
92 (FS) & $0.242\pm0.020$ | $0.137\pm0.022$
& $0.106\pm0.009$ | $0.060\pm0.010$ 
& $0.456\pm0.039$ | $0.260\pm0.042$ \\
97.5 (EK) & $0.219\pm0.010$ | $0.231\pm0.013$
& $0.086\pm0.004$ | $0.091\pm0.005$
& $0.370\pm0.016$ | $0.391\pm0.022$ \\
107.5 (EK) & $0.293\pm0.014$ | $0.279\pm0.019$ 
& $0.115\pm0.005$ | $0.109\pm0.007$ 
& $0.494\pm0.023$ | $0.473\pm0.032$ \\
117.5 (EK) & $0.311\pm0.012$ | $0.302\pm0.019$
& $0.122\pm0.005$ | $0.119\pm0.007$
& $0.525\pm0.020$ | $0.512\pm0.032$ \\
127.5 (EK) & $0.334\pm0.015$ | $0.328\pm0.025$
& $0.131\pm0.006$ | $0.129\pm0.010$
& $0.563\pm0.025$ | $0.555\pm0.042$ \\
152 (FS) &\dotfill | $0.255\pm0.046$
& \dotfill | $0.101\pm0.018$
& \dotfill | $0.437\pm0.079$ \\
158 (FS) &\dotfill | $0.291\pm0.031$
& \dotfill | $0.116\pm0.012$
& \dotfill | $0.499\pm0.053$ \\
164 (FS) &\dotfill | $0.248\pm0.030$
& \dotfill | $0.098\pm0.012$
& \dotfill | $0.424\pm0.051$ \\
170 (FS) &\dotfill | $0.264\pm0.025$ 
& \dotfill | $0.102\pm0.010$
& \dotfill | $0.441\pm0.042$ \\
176 (FS) &\dotfill | $0.267\pm0.022$ 
& \dotfill | $0.101\pm0.008$
& \dotfill | $0.436\pm0.036$ \\
181 (FS) &\dotfill | $0.286\pm0.025$
& \dotfill | $0.107\pm0.009$
& \dotfill | $0.463\pm0.041$ \\
186 (FS) &\dotfill | $0.324\pm0.027$ 
& \dotfill | $0.120\pm0.010$
& \dotfill | $0.518\pm0.043$ \\
195 (EK) & \dotfill |  $0.364\pm0.025$
& \dotfill | $0.143\pm0.010$
& \dotfill | $0.617\pm0.042$ \\
206 (FS) &\dotfill | $0.363\pm0.024$
& \dotfill | $0.135\pm0.009$ 
& \dotfill | $0.581\pm0.038$ \\
212 (FS) &\dotfill | $0.344\pm0.023$ 
& \dotfill | $0.127\pm0.009$
& \dotfill | $0.551\pm0.037$ \\
217 (FS) &\dotfill | $0.370\pm0.026$
& \dotfill | $0.134\pm0.009$
& \dotfill | $0.579\pm0.040$ \\
223 (FS) &\dotfill | $0.355\pm0.026$
& \dotfill | $0.129\pm0.009$
& \dotfill | $0.555\pm0.041$ \\
229 (FS) &\dotfill | $0.369\pm0.019$
& \dotfill | $0.134\pm0.007$
& \dotfill | $0.577\pm0.029$ \\
235 (FS) & $0.350\pm0.014$ | $0.380\pm0.017$ 
& $0.127\pm0.005$ | $0.138\pm0.006$
& $0.546\pm0.022$ | $0.138\pm0.006$ \\
246 (FS) & $0.279\pm0.013$ | $0.354\pm0.015$
& $0.103\pm0.005$ | $0.131\pm0.005$
& $0.444\pm0.021$ | $0.131\pm0.005$ \\
251 (FS) & $0.287\pm0.011$ | $0.368\pm0.013$
& $0.106\pm0.004$ | $0.136\pm0.005$ 
& $0.458\pm0.017$ | $0.136\pm0.005$ \\
257 (FS) & $0.315\pm0.010$ | $0.377\pm0.013$
& $0.118\pm0.004$ | $0.141\pm0.005$
& $0.507\pm0.017$ | $0.141\pm0.005$ \\
263 (FS) & $0.393\pm0.015$ | $0.398\pm0.020$ 
& $0.149\pm0.006$ | $0.151\pm0.008$
& $0.640\pm0.024$ | $0.151\pm0.008$ \\
269 (FS) & $0.315\pm0.013$ | $0.367\pm0.017$ 
& $0.119\pm0.005$ | $0.139\pm0.006$ 
& $0.513\pm0.020$ | $0.139\pm0.006$ \\
274 (FS) & $0.325\pm0.010$ | $0.366\pm0.014$
& $0.126\pm0.004$ | $0.142\pm0.005$
& $0.542\pm0.016$ | $0.142\pm0.005$ \\
280 (FS) & $0.357\pm0.019$ | $0.362\pm0.023$
& $0.138\pm0.007$ | $0.140\pm0.009$
& $0.559\pm0.031$ | $0.140\pm0.009$ \\
286 (FS) & $0.327\pm0.013$ | $0.359\pm0.019$
& $0.127\pm0.005$ | $0.139\pm0.007$
& $0.544\pm0.022$ | $0.139\pm0.007$ \\
292 (FS) & $0.328\pm0.018$ | $0.317\pm0.027$
& $0.130\pm0.007$ | $0.126\pm0.011$
& $0.559\pm0.031$ | $0.126\pm0.011$ \\
298 (FS) & $0.319\pm0.012$ | $0.361\pm0.017$
& $0.130\pm0.005$ | $0.147\pm0.007$
& $0.558\pm0.021$ | $0.147\pm0.007$ \\
304 (FS) & $0.276\pm0.013$ | $0.322\pm0.015$
& $0.115\pm0.005$ | $0.134\pm0.006$ 
& $0.494\pm0.023$ | $0.134\pm0.006$ \\
310 (FS) & $0.314\pm0.008$ | $0.330\pm0.012$ 
& $0.134\pm0.004$ | $0.141\pm0.005$
& $0.578\pm0.015$ | $0.141\pm0.005$ \\
317 (FS) & $0.297\pm0.015$ | $0.254\pm0.021$  
& $0.132\pm0.007$ | $0.113\pm0.009$
& $0.567\pm0.029$ | $0.113\pm0.009$ \\
325 (FS) &  $0.254\pm0.011$ | $0.261\pm0.018$ 
& $0.113\pm0.005$ | $0.116\pm0.008$  
& $0.485\pm0.021$ | $0.116\pm0.008$ \\
337 (FS) & $0.272\pm0.022$ | $0.249\pm0.031$
& $0.123\pm0.010$ | $0.112\pm0.014$
& $0.528\pm0.043$ | $0.112\pm0.014$\\
\hline
Mean &  0.302 (0.059)
&  0.121 (0.022)
&  0.522 (0.094)
\enddata

\tablecomments{Errors quoted are statistical 1-sigma uncertainties.
  Systematic uncertainties for proper motions, expansion rates, and
  expansion indices are 0$^{\prime\prime}$.061\,yr$^{-1}$ (or
  0$^{\prime\prime}$.072\,yr$^{-1}$), $\sim$0.024\,\%\,yr$^{-1}$ (or
  $\sim$0.028\,\%\,yr$^{-1}$), and $\sim$0.103 (or $\sim$0.122) for
  2000-2007 (or 2003-2007), respectively.  The values in blackets with
  mean values are standard deviations.  
}
\label{tab:results1}
\end{deluxetable}

\begin{deluxetable}{lccccc}
\tabletypesize{\tiny}

\tablecaption{Summary of proper-motion measurements for reverse-shocked ejecta}
\tablewidth{0pt}
\tablehead{
\colhead{Sector}&\colhead{Proper
  motion (arcsec\,yr$^{-1}$)}&\colhead{Expansion rate
  (\%\,yr$^{-1}$)}&\colhead{Expansion index}  
}
\startdata
&2000-2007 | 2003-2007&2000-2007 | 2003-2007&2000-2007 | 2003-2007
\\
\hline
NE(0$^\circ$--60$^\circ$) & $0.210\pm0.005$ | $0.222\pm0.007$
& $0.100\pm0.002$ | $0.106\pm0.003$
& $0.429\pm0.009$ | $0.457\pm0.014$ \\
SE(132$^\circ$--180$^\circ$) & \dotfill | $0.266\pm0.013$
& \dotfill | $0.133\pm0.007$ 
& \dotfill | $0.574\pm0.029$ \\
SW(180$^\circ$--240$^\circ$) & \dotfill | $0.297\pm0.008$
& \dotfill | $0.145\pm0.004$ 
& \dotfill | $0.625\pm0.017$ \\
W(240$^\circ$--300$^\circ$) & $0.291\pm0.004$ | $0.313\pm0.006$
& $0.138\pm0.002$ | $0.149\pm0.003$
& $0.596\pm0.008$ | $0.644\pm0.013$ \\
NW(300$^\circ$--360$^\circ$) & $0.221\pm0.004$ | $0.237\pm0.007$
& $0.106\pm0.003$ | $0.118\pm0.003$
& $0.473\pm0.008$ | $0.509\pm0.014$ \\
\hline
Mean &  0.294 (0.040)
& 0.143 (0.019)
& 0.614 (0.084)
\enddata

\tablecomments{Same as table~\ref{tab:results1}. 
}
\label{tab:results2}
\end{deluxetable}

\end{document}